\documentclass[twocolumn,showpacs,preprintnumbers,amsmath,amssymb]{revtex4}

\usepackage{graphicx}
\usepackage{dcolumn}
\usepackage{bm}

\newcommand{\s}{{\sigma}}

\newcommand{\w}{{\omega}}

\newcommand{\be}{\begin{eqnarray}}
\newcommand{\ee}{\end{eqnarray}}
\newcommand{\nn}{\nonumber\\}
\newcommand{\Eq}[1]{Eq.~(\ref{#1})}
\newcommand{\p}{\partial}

\newcommand{\ra}{\rightarrow}

\begin{document}
\draft
\title{Role of interference in MM-wave driven DC transport in two
dimensional electron gas}
\author{Dung-Hai Lee$^{a,b}$, and Jon Magne Leinaas$^{a,c}$}
\affiliation{${(a)}$Department of Physics,University of California
at Berkeley, Berkeley, CA 94720, USA}\affiliation{${(b)}$ Material
Science Division, Lawrence Berkeley National
Laboratory}\affiliation{${(c)}$Department of Physics,University of
Oslo, P.O. Box 1048 Blindern, 0316 Oslo, Norway}

\begin{abstract}
In this paper we point out that in addition to the density of
states effect proposed in Ref.\cite{durst,anderson} one should
consider the effect of constructive interference between the
multi-MM-wave-photon processes shown in Fig.2. This process
enhances the dark value of the conductivity. When the sample is
very pure, i.e., when the transport life time is very long, this
interference effect quickly diminishes as the MM-wave frequency
deviates from the cyclotron frequency. In this paper we also
present the linear response theory in the presence of strong
harmonic time-dependent perturbation.
\end{abstract}
\maketitle \draft

  \narrowtext

The recent observation of strong suppression of the longitudinal
resistivity of a two-dimensional electron gas (2DEG) by a
millimeter (MM) radiation source \cite{zudov,mani} has stimulated
considerable interests in the condensed matter
community
\cite{durst,anderson,millis,Shi,Phillips,Begeret,
Dorozhkin,Ryzhii,Shrivastava,Rivera}. In Ref.
\cite{durst} and Ref.\cite{anderson} it is pointed out that the
combined effect of photo-excitation by the MM wave photon and
scattering by impurities can lead to a sinusoidal modulation of
the conductivity as a function of $\w_0/\w_c$ (the MM wave photon
frequency over the cyclotron frequency). When the amplitude of
this modulation becomes big, the conductivity becomes negative and
the system become unstable. In Ref.\cite{anderson} and
Ref.\cite{millis} it is postulated that this leads the system to
self-organize into a state with zero conductivity.

In this work we propose another mechanism that may also be of
importance for the observed phenomenon. In essence our mechanism
associate the above experimental results with a phenomenon called
``electromagnetically induced transparency'' \cite{eit}. This
phenomenon occurs when an optical transition can take place
through many alternative processes. When these processes
destructively interfere with one another, the net optical
transition amplitude vanishes. This has been observed when two of
the levels of a three-level system is resonantly coupled together
by a strong coupling laser. As a consequence, the resonant peak
due to the absorption from the third level to one of these
strongly mixed states is suppressed to zero.

In the present case we assume that the MM wave is sufficiently
strong to couple many electronic states together. When an
additional low frequency probing source is turned on, the
transition that involves the absorption/emission of a single
probing photon can occur via many virtual processes with varying
number of MM wave photons absorbed and subsequently emitted. The
interference between these processes can lead to
electromagnetically induced transparency. The effect is
illustrated in Fig.1 where the vertical arrows indicate the
emission/absorption of the MM wave photons, and the slanted
horizontal arrows indicate the absorption of the probing photons.
If the matrix elements associated with the slanted horizontal
arrows at different vertical levels are in phase, this leads to
constructive interference. Otherwise destructive interference will
be resulted. In the DC limit (vanishing probing photon frequency)
destructive interference implies a strong suppression of the
longitudinal conductivity. This is equivalent to a suppression of
the longitudinal resistivity when $\s_{xy}>>\s_{xx}$. In the
following we shall argue that the ratio between the MM wave
frequency and the cyclotron frequency determines whether the
interference is constructive or destructive.

\begin{figure}
\includegraphics[width=7.2cm,height=3.3cm]{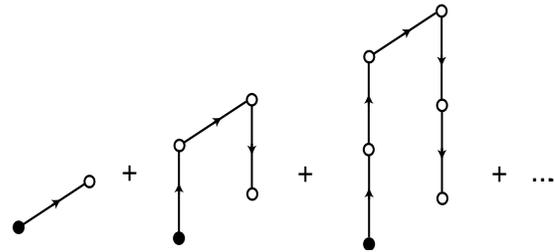}
\caption{The processes in this figure and their particle-hole
inversion analog causes interference discussed in the text. The
vertical lines represent absorption or emission of the coupling
field photons. The slanted horizontal arrows correspond to the
absorption of the probing field photon. }
\end{figure}

Our picture of this $\w_0/\w_c$ dependence is the following. When
$\w_0$ is an integral multiple of $\w_c$, the MM wave couples
together states that have (essentially) the same guiding center
orbit but different Landau level indices. Due to energy
conservation we argue that the absorption/emission of a (low
frequency) probing photon can only change the guiding center
orbits not the Landau level index. More explicitly we assume that,
with the initial and final guiding center orbits of the processes
depicted in Fig.1 being all the same, the matrix elements
corresponding to the slanted horizontal arrows are approximately
equal. The processes shown in Fig.1 tend therefore to interfere
constructively. However, when $\w_0$ is not close to an integer
multiple of $\w_c$, the initial and final guiding center orbits of
the different slanted horizontal arrows are  different. This will
generally give rise to matrix elements with different phases and
lead to destructive interference.

The mechanism discussed here (like in \cite{durst,anderson}) the
electron-electron interaction does not play a central role.
However it is important to keep in mind that electron-electron
interaction affects the  self-consistent potential each electron
sees hence affect the eigenstates $|i>$ in \Eq{repl}. In
particular in the mechanism proposed in Ref.\cite{millis} where
significant redistribution of charges can be resulted by the
action of the MM wave radiation, the steady-state self-consistent
potential can differ significantly from that in the absence of MM
wave radiation. In the rest of the paper we shall ignore the
electron-electron interaction with the understanding that the
potential the electrons see is the steady-state self-consistent
one.

In the following we first develop the formalism for describing
this problem. We obtain a Kubo-like formula (\Eq{kubo}) that
allows one to compute the AC/DC conductivity in the presence of a
strong harmonic time-dependent driving field. We would like to
stress that method developed here is not restricted to the
specific problem considered. Rather it is a general formulation of
linear response in the presence of a strong time-dependent
perturbation, and in this respect it is a step in the direction of
making a systematic approach to describing the behavior of a
many-particle system in a presence of a strong oscillating field.

We emphasize that the mechanism described here is requires the
assumption that MM wave field acts {\it coherently} on the system.
It is of interest to look for experimental tests of the importance
of coherence. We suggest that application of two radiation sources
to the same sample (see the concluding paragraph) is a good way to
determine  how important the interference effects are.

In the rest of the paper we shall use the term {\it coupling
field} to denote the MM radiation field, and {\it probing field}
to denote the low frequency field associated the linear response
measurement.

\section{The general formalism}

\subsection{The Floquet eigenvalue problem}

Let us assume $\{|i>\}$ are the exact {\it single particle}
eigenstates (in the presence of disorder) in the absence of the
coupling field
\begin{equation}H_0 |i> = \epsilon_i |i>\label{repl}.\end{equation} With the
coupling field turned on, the Hamiltonian becomes \be H(t) &&= H_0
+ g_c \Big[ e^{-i\w_0t} \sum_{ij} P_{ij} |i><j|\nn&&+
e^{i\w_0t}\sum_{ij} P_{ji}^*|i><j|\Big].\label{hm}\ee In the above
$\w_0$ is the frequency of the coupling field, and $P_{ij}$ is the
matrix element of the current operator.

The solution of the time-dependent Schr\"{o}dinger equation
\begin{equation}i\p_t|\psi(t)>=H(t)|\psi(t)>\end{equation}can be written as
\begin{equation}|\psi(t)> = \sum_i \phi_i(t) |i>,\label{ghh}\end{equation}where
$\phi_i(t)$ satisfies \be i\p_t \phi_i(t) &&= \epsilon_i \phi_i(t)
+ g_c e^{-i\w_0 t }\sum_j P_{ij}\phi_j(t) \nn&&+ g_c e^{i\w_0 t
}\sum_j P^*_{ji}\phi_j(t).\label{fq}\ee Due to the symmetry of
\Eq{fq} upon time translation
\begin{equation}t\ra t+n{2\pi\over\w_0},\end{equation} the solutions of
of \Eq{fq} $\phi_i^{\alpha}(t)$ can be written in the form
\begin{equation}\phi_i^{\alpha}(t)=e^{-iE_{\alpha} t}\sum_n
\phi_{ni}^{\alpha} e^{-i
n\w_0t}.\label{bkk}\end{equation}
where the index $\alpha$ labels the different solutions. In the above
\begin{equation}-{\w_0\over 2}\le E_{\alpha}<{\w_0\over
2}\label{brill}\end{equation}is the ``Brillouin zone'' in
frequency. This time version of the Bloch theorem is called the
Floquet theorem.\cite{fq}. By substituting \Eq{bkk} into \Eq{fq}
and equate coefficients of $e^{-i(E_{\alpha}+n\w_0)t}$ we obtain
the time independent equation \be E_{\alpha} \phi_{ni}^{\alpha}=
(\epsilon_i-n\w_0)\phi_{ni}^{\alpha} + g_c \sum_j
\Big[P_{ij}\phi_{n-1j}^{\alpha} +
P^*_{ji}\phi_{n+1j}^{\alpha}\Big].\label{eig}\ee \Eq{eig} is an
algebraic eigenvalue problem which can be solved to obtain
eigenvalue $E_{\alpha}$ and corresponding eigenvector
$\phi_{ni}^{\alpha}$.  Since $\phi^{\alpha}_{ni}$ and
$\phi^{\beta}_{ni}$ are eigenvectors of \Eq{eig}, they satisfy
(when properly normalized) the orthogonality relation
\begin{equation}\sum_{n,i}({\phi}^{\beta}_{ni})^*\phi^{\alpha}_{ni}
=\delta_{\alpha\beta}.\label{or}\end{equation}

It is straightforward to prove the following fact concerning the
solutions of \Eq{eig}. If $\phi^{\alpha}_{ni}$ is the
eigenvector of \Eq{eig} with eigenvalue $E_{\alpha}$, then
\begin{equation}\phi^{\alpha'}_{ni}\equiv \phi^{\alpha}_{n+1
i}\label{trsf}\end{equation}is also an eigenvector of \Eq{eig}.
The eigenvalue associated with the latter is
\begin{equation}E_{\alpha'}= E_{\alpha}+\w_0.\label{trsf2}\end{equation}
By repeated application of \Eq{trsf} we can generated a family of
solutions
$\phi^{\alpha}_{ni},\phi^{\alpha'}_{ni},\phi^{\alpha''}_{ni}...$
of \Eq{eig}. It is simple to show that all these solutions lead to
the same time-dependent solution \be
|{\bar{\alpha}}(t)>=e^{-iE_{\alpha}t}\sum_{in}\phi^{\alpha}_{ni}e^{-in\w_0
t}|i>.\label{pol}\ee In the above $\bar{\alpha}$ denotes the
entire class of Floquet eigenvectors related by \Eq{trsf} and
\Eq{trsf2}.

It is simple to prove that
\begin{equation}<{\bar{\beta}}(0)|{\bar{\alpha}}(0)>
=\delta_{\bar{\alpha}\bar{\beta}}.\label{rt}\end{equation} Thus
\begin{equation}|\bar{\alpha}>\equiv
|{\bar{\alpha}}(0)>=\sum_{ni}\phi^{\alpha}_{ni}|i>\label{def}\end{equation} can
be used as an orthonormal basis just as $\{|i>\}$.
We also note that because the time evolution is unitary, \Eq{rt}
ensures that
\begin{equation}<{\bar{\beta}}(t)|{\bar{\alpha}}(t)>
=\delta_{\bar{\beta}\bar{\alpha}}\end{equation} for any $t>0$.

Now we have obtained the an orthonormal set of solution of the
time-dependent Schr\"{o}dinger equation
\begin{equation}\{|{\bar{\alpha}}(t)>\}.\end{equation}Of course, any
linear combination of these solutions
\begin{equation}\sum_{\bar{\alpha}}A_{\bar{\alpha}}|{\bar{\alpha}}(t)>\label{ini}\end{equation}is
itself a solution of the Schr\"{o}dinger equation. The coefficient
$A_{\bar{\alpha}}$ in \Eq{ini} are determined by the initial condition. For
example by properly choosing
$A_{\bar{\alpha}}$ we can construct a orthonormal set of solutions
$\{|i(t)>\}$ satisfying the Schr\"{o}dinger equation and the
initial condition that
\begin{equation}|i(0)>=|i>.\end{equation}Using \Eq{rt} and
\Eq{def} it is simple to show that
\begin{equation}|i(t)>=\sum_{\bar{\alpha}} <\bar{\alpha}|i>
|{\bar{\alpha}}(t)>.\end{equation}


To gain some intuitive feeling for what \Eq{eig} represents we
consider the following model. For each $n$ in \Eq{eig} we define a
replica of the physical system with eigenspectrum given by
\Eq{repl}. These replicas are coupled together by $P_{ij}$ in the
following Hamiltonian \be H_r&&=\sum_n\sum_{i}(\epsilon_i-n \w_0)
c^+_{ni}c_{ni}+g_c\sum_n\sum_{ij}[P_{ij}c^+_{ni}c_{n-1j}\nn&&+h.c.].\label{3d}\ee
In the above $c^+_{ni}$ create an electron in the ith states of
the nth replica. The physical interpretation of the $n$ variable
is the photon number of the coupling field. The single particle
states in the nth layer are $\{|n,i>\}$, and the associated eigen
energies are $\epsilon_i$. The replicas are coupled together by
hopping (the terms proportional to $g_c$), and an ``electric
field'' is turned on so that
the potential energy of the nth replica is $-n\w_0$. 
The replica model is constructed so that \Eq{eig} is its
time-independent eigen equation. Each solution
$\phi^\alpha_{ni}$ of \Eq{eig} uniquely defines an eigen state
\begin{equation}
\sum_{n,i} \phi^{\alpha}_{n,i} |n,i>\label{ewr}\end{equation} of
the replica model. \Eq{trsf} and \Eq{trsf2} link a whole family of
replica states together. 

\subsection{The linear response theory}

In this section we derive the formula for the AC conductivity in
the presence of the coupling field. The formalism developed in
this section is rather general. The only restriction is that the
electron-electron interaction is neglected.

The AC conductivity is the response of the system to a
time-dependent probing field. The Hamiltonian in the presence of
such probing field is given by \be
&&H'(t)=H(t)+H_p(t)\nn&&H_p(t)=g_p e^{-i\w_pt} \sum_{ij} D_{ij}
|i><j|+ h.c., \label{hm1}\ee
where $D_{ij}$ is the current matrix element coupling to the
probing field. Unlike the coupling field, the probing field is
{\it very weak}. Therefore we will treat its effect perturbatively
to the lower order in $g_p$.

If we use $|i(t)>$ as basis, the probability amplitude that the
probing field will induce a transition from $|i>$ at time zero to
$|j(t)>$ at time t is given by
\begin{equation}{A}_{ji}=\delta_{ji}-i \int_0^t d\tau
<j(\tau)|H_p(\tau)|i(\tau)>+...\end{equation} If the we model the
decoherence by a single decoherence time $\tau=1/\Gamma$ the
the averaged transition rate between $i\ne j$ is given by \be
W_{ji}=
2\Gamma~\Big{|}\int_0^{\infty}dt e^{-\Gamma t
}<j(t)|H_p(t)|i(t)>\Big{|}^2.\ee Straightforward calculation shows
that
\begin{equation}W_{ji}=W^a_{ji}+W^e_{ji}\end{equation}(``a'' stands for
absorption and ``e'' stands for emission), where \be &&W^a_{ji}=
2\Gamma
\Big{|}\sum_l\sum_{{\alpha}{\beta}}{<j|\bar{\alpha}>D^l_{{\alpha}{\beta}}<\bar{\beta}|i>\over
E_{{\alpha}}-E_{{\beta}}-\w_p+l\w_0+i\Gamma}\Big{|}^2\nn
&&W^e_{ji}= 2\Gamma
\Big{|}\sum_l\sum_{{\alpha}{\beta}}{<j|\bar{\alpha}>\Big(D^{l}_{{\beta}{\alpha}}\Big)^*<{\beta}|i>\over
E_{{\alpha}}-E_{{\beta}}+\w_p+l\w_0+i\Gamma}\Big{|}^2.\label{rate}\ee
In \Eq{rate} \be
D^l_{{\alpha}{\beta}}\equiv\sum_n\sum_{ij}({\phi}^{\alpha}_{n+li})^*D_{ij}\phi^{\beta}_{nj}.\label{D}\ee
For the $\alpha$ and $\beta$ in the summation we choose one
$\alpha$ and one $\beta$ for each family of solutions $\bar\alpha$
and $\bar\beta$.  We note that if these are chosen so that
$E_{{\alpha}}-E_{{\beta}}$ is small compared to $\omega_0$ (we
shall assume this choice in the rest of the paper), the main
contribution in the sum over $l$ comes from $l=0$, since
$\omega_p$ is small.

The total absorption and emission rates and the AC conductivity
are given by \be &&{\cal{A}}=\sum_{ij}f_i(1-f_j)~W^a_{ji}\nn
&&{\cal{E}}
=\sum_{ij}f_i(1-f_j)~W^e_{ji}\nn&&\sigma_{xx}(\w_p)\sim {1\over
{\rm \Omega}}\lim_{\w_p\ra 0}{1\over
\w_p}({\cal{A}}-{\cal{E}}),\label{kubo}\ee where
$f_i=f(\epsilon_i)$ is the Fermi function and $\Omega$ is the
volume (or area for 2D) of the sample.

In the limit of $\Gamma\ra 0$ only the diagonal terms in the
expansion of $|...|^2$ in \Eq{rate} survive. In that case
$E_{\alpha}-E_\beta\pm\w_p$ is constrained to zero.
For finite $\Gamma$ on the other hand $E_{\alpha}-E_{\beta}$ is
not strictly fixed by energy conservation, but can take value
within a range $\Gamma$ around $\pm \omega_p$. In this case
${\cal{A}}$ and ${\cal{E}}$ do not only get contributions from the
diagonal terms in the expansion of (\ref{rate}) but also from
crossing terms. When the phase coherence length is much smaller
than the sample dimension, the above formula should also be
averaged over the configurations of the random potential.

\Eq{rate}, \Eq{D} and \Eq{kubo}, which express the linear response
coefficients of a strongly driven system in terms of the
eigenenergies and eigenfunctions of a time-independent Hamiltonian
($H_r$ in \Eq{3d}), is a main result of this paper.




\subsection{A simple model}
In order to illustrate the application of the above formalism we
now consider a simple model where the ``dark'' energy levels are
labeled by two parameters, $i=(\nu, a)$, and the energies are
given by \be \epsilon_{\nu a}=\nu \omega_0 +V_a \ee where
$-\w_0/2\le V_a<\w_0/2.$ In the following we shall refer to $\nu$
as the ``vertical'' and $a$ as the ``horizontal'' indices. 
We shall assume resonant coupling, i.e., the absorption/emission
of a coupling photon results in vertical transition $|\nu,a>\ra
|\nu\pm 1,a>$, and the the absorption/emission of a probing photon
results in horizontal transition $|\nu,a>\ra |\nu,b>$.
More specifically we consider the following Hamiltonian
\be H(t)&&=H_0+H_c(t)+H_p(t)\nn H_c(t)&&=g_c \Big[ e^{-i\w_0t}
\sum_{\nu a} P^\nu_{a} |\nu+1, a><\nu, a| + h.c.\Big]\nn H_p(t)
&&= g_p\Big[e^{-i\w_p t }\sum_{\nu ab}D_{ab}^{\nu}|\nu, a><\nu, b|
+ h.c.\Big].\label{model1}\ee  To simplify the Floquet eigen equation we
further assume \be P^\nu_{a}=P.\label{simp}\ee 
Given \Eq{model1} and \Eq{simp} the Floquet eigenvectors are given
by \be \phi^{\alpha}_{ni}=\phi^{\alpha}_n
\delta_{\nu,\nu_{\alpha}+n}\delta_{a,a_{\alpha}} \ee where
$\nu_{\alpha}$ and $a_{\alpha}$ are parameters that characterize
the Floquet eigenstate $\alpha$, and  $\phi^{\alpha}_n$ satisfies
\be
E_{\alpha}\phi^{\alpha}_n=(V_{a_{\alpha}}+\nu_{\alpha}\omega_0)\phi^{\alpha}_n
+ g_c(P\phi^{\alpha}_{n-1}+P^*\phi^{\alpha}_{n+1}) \ee This
equation is invariant under translations $n\rightarrow n+1$ and
has, according to Bloch's theorem, solutions of the form \be
\phi^{\alpha}_n={1\over\sqrt{N}}e^{in\theta_{\alpha}}\ee For
simplicity we assume the variable $n$ to take a finite number of
values $N$, with $\phi^{\alpha}_n$ as a periodic function. This
implies \be \theta_{\alpha}=\frac{2 \pi}{N}n_{\alpha},
\;\;\;n_{\alpha}=0,1,...,N-1. \ee with $n_{\alpha}$ as a new
discrete parameter that characterizes the state $\alpha$. The
energy of the Floquet state $\alpha$ now is given by \be
E_{\alpha}=\nu_{\alpha}\omega_0 +
\lambda\cos(2\pi\frac{n_{\alpha}}{N}+\delta) \ee with the
parameters $\lambda$ and $\delta$ defined by \be \lambda
e^{-i\delta}=g_cP. \ee

With the expressions for $\phi^{\alpha}_{n i}$ and $E_{\alpha}$
inserted in \Eq{rate}, the absorption/emission rate gets the form
\be &&W^{a/e}_{ji}=2\Gamma \frac{1}{N^2}\times\nn&& \Big{|}\sum_{p
q}{{1\over N}\sum_n e^{{2\pi i}\frac{p}{N} n} D^{\nu+n}_{a b}
\over V_{ab}+2\lambda\sin(2\pi{p \over N})sin(2\pi{q \over
N}+2\delta)\pm\w_p+i\Gamma}\Big{|}^2.\nn \label{rate2}\ee with
$V_{ab}=V_a-V_b$. This expression is illustrated by the diagram of
Fig.1, referred to in the introduction. Without the coupling field
turned on the transition induced by the probing field is
restricted to one value of $\nu$ for each pair $(a,b)$. However,
with the coupling field on, the contributions from higher levels
$\nu+n$ are introduced, as illustrated by the vertical arrows in
the diagram and by the sum over $n$ in the formula.

The variations of $D^{\nu+n}_{a b}$ with $n$ may give rise to
terms in \Eq{rate2} that adds constructively or destructively. If
we assume no dependence on $n$, $D^{\nu+n}_{a b}=D_{a b}$, the sum
over $n$ can be explicitly done giving rise to $N\delta_p$, hence
$p$ only gets contribution from $p=0$ and the expression reduces
to its "dark value", {\em i.e.} its value without the coupling
field \be &&W^{a/e}_{ji}= 2\Gamma \Big{|} { D_{a b} \over
V_{ab}\pm\w_p+i\Gamma}\Big{|}^2. \label{dark}\ee However, if the
phase of  $D^{\nu+n}_{a b}$ varies randomly with $n$ \be
\Big{|}{1\over N}\sum_n e^{{2\pi i}\frac{p}{N} n} D^{\nu+n}_{a
b}\Big{|}\sim{1\over\sqrt{N}}\label{ineq}\ee and the phase
generically varies randomly with $p$. In the limit $\Gamma\ra 0$
only the diagonal terms in the expansion of $|...|$ in \Eq{rate2}
contribute and, owing to \Eq{ineq}, $W^{a/e}_{ji}\sim
N^2/N^3=1/N$. Hence we have destructive interference. When
$\Gamma\ne 0$ but much smaller compared to $\lambda$, the
non-diagonal terms become important. In that case simple estimate
leads to \be W^{a/e}_{ji}\approx{K\over\Gamma}\Big[{c_1\over
N}+{c_2\over\sqrt{N}}{\Gamma\over\lambda}+c_3\Big({\Gamma\over\lambda}\Big)^2\Big]\label{des}
,\ee where $K,c_1,c_2,c_3$ are $n$-independent constants. From
\Eq{des} we see that the suppression by large $N$ depends on
$\Gamma<<\lambda$. It is important to note that in the simple
model considered here the suppression of the DC/AC conductivity
is complete only when $N\ra\infty$ and $\Gamma/\lambda\ra 0$.


In the model we have introduced several simplifications, in
particular only including resonant couplings. However, we believe
this model convey the essence of our idea. We note however, that
with the simplifications introduced we do not see enhancement of
the amplitudes for constructive interference, but rather
  a return to the dark value.

\section{Application to the 2-dimensional electron gas.}

When discussing the application of the general formalism to the
2DEG, there are two regimes of interest to discuss separately. The
first one is the weak coupling limit, where the Floquet state is
well localized with respect to layer index $n$ and where the
coupling field as well, as the probing field, can be treated
perturbatively. The other case is the strong coupling regime,
where the state is extended through many $n$'s. We first treat the
weak coupling case and discuss this in a general way with focus
especially on the effect of the density of states. The strong
coupling case we discuss more qualitatively, with specific
reference to the simple model discussed in the previous section.

\subsection{Weak coupling. The density of states effect.}

In this limit the eigenstates of $H_{r}$ are qualitatively similar
to those in the absence of the hopping between the replica's. In
this limit the our calculation produces result similar that
obtained in Ref.\cite{durst} and Ref.\cite{millis}. For a
comparison with these references we restrict ourself to the case
of total coherence, i.e. $\Gamma\ra 0$.

Assuming $\phi_{ni}^{\alpha}$ leaks weakly to the adjacent layers
we obtain (upon using \Eq{def}) \be &&|\bar{\alpha}
>\approx |j> + \sum_{j'} \eta_{jj'} |j'> + \sum_{j''}
\chi_{jj''}|j''>\nn&&|\bar{\beta}
>\approx |i> + \sum_{i'} \eta_{ii'} |i'> + \sum_{i''}
\chi_{ii''}|i''>.\label{gh}\ee In the above $|i'>,|j'>$ are states
with energy $\w_0$ above those of of $|i>,|j>$ and $|i''>,|j''>$
has energy $\w_0$ below those of $|i>,|j>$. In the rest of this
section we shall treat $\eta$ and $\chi$ as first order in the
coupling constant $g_c$ and obtain ${\cal{A}}$ and ${\cal{E}}$ to
$O(g_c^2)$ by substituting \Eq{gh} into \Eq{kubo} and take the
$\Gamma\ra 0$ limit. The result for ${\cal{A}}$ is


\be {\cal{A}}&&=2\pi g_p^2\Big{\{}
\sum_{ij}\delta(E_{ji}-\w_p)f_i(1-f_j)|D_{ji}+\delta D_{ji}|^2
\nn&&+\sum_{ij}\delta(E_{ji}-\w_0-\w_p)f_i(1-f_{j}) |\delta
D'_{ji}|^2\nn&&+\sum_{ij}\delta(E_{ji}-\w_p)f_i\Big[\sum_{j'}(1-f_{j'})|\eta_{jj'}|^2\Big]|D_{ji}|^2\nn&&+
\sum_{ij}\delta(E_{ji}-\w_p)\Big[\sum_{i''}|\chi_{ii''}|^2f_{i''}\Big](1-f_j)|D_{ji}|^2
\Big{\}}.\nn&&\label{hww}\ee In the above
\begin{equation}E_{ji}\equiv E_j-E_i,\end{equation}and $\delta D_{ji}$,
$\delta D'_{ji}$ are of order $O(\epsilon)$ respectively.

The first term of \Eq{hww} is the value in the absence of coupling
field(the matrix elements are slighted modified). 
distribution function due to the coupling laser. 
If we assume that the matrix elements in \Eq{hww} are smooth
function of energy we can replace this equation by \be
&&{\cal{A}}=2\pi g_p^2\Big[\int dE
f(E)(1-f(E+\w_p))N(E)\nn&&\times N(E+\w_p)|M_1|^2+\int dE
N(E)N(E+\w_0+\w_p)\nn&& \times
f(E)(1-f(E+\w_0+\w_p))|M_2|^2\Big],\label{hww1}\ee where $M_1$ and
$M_2$ are $E$-dependent functions that we do not write out
explicitely. The corresponding emission term is given by \be
&&{\cal{E}}=2\pi g_p^2\Big[\int dE
f(E)(1-f(E-\w_p))N(E)\nn&&\times N(E-\w_p)|M_1|^2+\int dE
N(E)N(E+\w_0-\w_p)\nn&& \times
f(E)(1-f(E+\w_0-\w_p))|M_2|^2\Big]\label{hww2}\ee From \Eq{hww1}
and \Eq{hww2} we can compute the DC conductivity via
\begin{equation}\sigma_{xx}\sim\lim_{\w_p\ra
0}{{\cal{A}}-{\cal{E}}\over\w_p}.\label{fogrm}\end{equation}

If we take a simple form
\begin{equation}N(E)=N_0+N_1\cos\Big({2\pi E\over
\w_c}\Big),\end{equation}as suggested at the beginning of
Ref.\cite{durst}, and assume $M_1$ and $M_2$ to be functions
changing slowly with $E$, we obtain from \Eq{fogrm}, \Eq{hww1} and
\Eq{hww2} the result
\begin{equation}\rho_{xx}=\rho_{xx}^0 -\rho^1_{xx}\sin\Big({2\pi
\w_0\over \w_c}+\phi\Big),\end{equation}where
$\phi=\tan^{-1}(k\w_c/\pi\w_0)$ and $k=1+2N_0/N_1$ . This result
is in qualitative agreement with that obtained in
Ref.\cite{durst,anderson}.


\subsection{Strong coupling. Constructive and destructive interference.}

When the coupling constant $g_c$ in \Eq{hm} is large, the
eigenstates of \Eq{3d} become extended among many replicas, {\em
i.e.}
$\phi^{\alpha}_{ni}$ is non-zero for a very wide range of $n$. %
For the electrons in the magnetic field this means that the MM field
couples together many states, generally located at different guiding center
orbits. The states coupled together will be selected by the condition of
having  the energy difference in resonance (or close to resonance) with the
oscillating field.

The dipole matrix element of the coupling laser primarily couples
states at neighboring Landau levels and the same guiding center
orbit, or close by guiding center orbits in the same Landau level.
However, the observed effect, with strong variations in the
conductivity when the frequency of the coupling field matches
integer multiples of the cyclotron frequency, indicates a rather
strong coupling also between non-adjacent Landau levels. Such
coupling clearly requires Landau level mixing. This mixing may be
due to the presence of a dilute concentration of  strong and
localized scatterers and/or high gradients in the
potential at the sample edge. 
In this picture the electrons mainly drift adiabatically in a
smooth background potential, but occasionally encounter regions of
high potential gradients where the Landau levels is mixed. Due to
the dilute concentration of such regions, the effect on e.g. the
density of state can be very small. However for a strong coupling
field this is enough to generate sufficient coupling to distant
Landau levels/distant guiding center orbits that is required to
delocalize the Floquet eigenstates among many replicas. For the
the transitions induced by the probing field this mixing is less
important, since the low frequency transitions are limited to
close by orbits in the same Landau level.

Let us first consider the case where the coupling source is tuned
to one of the frequencies $m \omega_c$, with integer $m$. The
matrix elements $D_{ij}$ of the probing field in the sum of \Eq{D}
now are defined between the {\it same} pair of guiding center
orbits elevated to higher Landau levels by the coupling field. The
model of the previous section can be viewed as giving a simplified
description of this situation, with the "vertical" index $\nu$
identified with the Landau level index and the "horizontal" index
$a$ labelling the guiding center orbit.

The matrix elements $D^{\nu}_{ab}$ now describe transitions
between the same two guiding center orbits $a$ and $b$ at
different Landau levels $\nu$. These matrix elements are strongly
correlated and we have checked that in an adiabatic approach they
are only slowly dependent on the Landau level index $\nu$.

As previously discussed, with $D^{\nu}_{ab}$ independent of $\nu$,
the sum over $n$ in \Eq{rate2} will include terms that interfere
constructively.  
Although one should note
that in the expression found there was no enhancement relative to
the dark value. Such an enhancement is clearly present in the
observed effect. We believe such an enhancement is due to an
increase in the {\it magnitude} of $D^{\nu}_{ab}$ with $\nu$
and/or to the effect of including non-resonant terms both of which
are ignored in our simple model.

Let us next turn to the case where the coupling frequency is not
close to $n \omega_c$. 
In this case different $|\nu,a>$ and $|\nu+1,a>$ label two distant
guiding center orbits with the potential energy of the second
orbit $\w_0$
higher than that of the first.

Since the matrix element $D^{\nu}_{ab}$, for different $\nu$,
refers to transitions between distant pairs of guiding center
orbits, they are no longer strongly correlated and may change
substantially from one value of $\nu$ to the next. In an adiabatic
approximation the matrix element can be related to variations in
the drift velocity along the guiding center orbit, and in a
randomly varying background potential we therefore expect a
corresponding random variation in the matrix element.  In this
case the $n$ sum in \Eq{rate2} add destructively.

To recap, the difference between the case $\w_0=m\w_c$  and
$\w_0\ne m\w_c$ lies in the fact that for the former the $n$ sum
in the absorption/emission amplitude tend to add
non-destructively, whereas for the latter the sum tend to add
destructively. Thus, we propose that in addition to the density of
state mechanism discussed in Ref.\cite{durst,anderson}, and seen
in our discussion of the weak coupling limit, the above
interference mechanism will contribute, for $\w_0$  between
integer multiples of $\w_c$, to suppress the {\it magnitude} of
the longitudinal resistivity.  This can reduce the resistivity or
even suppress it to zero when the effect is very strong. However,
since our discussion at this point is qualitative, we cannot
estimate the real strength of the effect.

\section{Summary}

In this paper we present a general formalism for describing the
DC/AC transport in the presence of an oscillating coupling field.
The oscillating field couples together states in the form of a
Floquet state, and by use of the expression for this state we
derive a general form for the absorption and emission
probabilities of an additional low frequency probing field. This
gives a generalized Kubo formula for the conductivity in the
presence of an oscillating field.

We further discuss in a general way the application of this to the 2D
electron gas in a magnetic field radiated by a MM wave. For a weak coupling
field we show the presence of oscillations due to variations
in the density of states. For strong coupling we discuss in a qualitative
way the difference between constructive interference when the frequency
matches the energy difference between two Landau levels and destructive
interference for intermediate values of the frequency. The discussion of
the interference effects is illustrated by a simplified model.

Quantum interference between different virtual processes described
in this paper should in general exist. However, complete
destructive interference requires a large number of interfering
processes and small decoherence. We are currently uncertain about
the values of these parameters for the experimental system. The
mechanism presented here raises the question of whether an
independent study of the strength of coherence effects can be
performed. Clearly, if the coherent MM source is replaced by an
incoherent source at the same frequency the coherence effects will
be destroyed. As a simpler experiment we suggest that the
importance of coherence for the observed effect can be studied by
use of two independent coupling fields. If one of these is tuned
to one of the peak values (say $2\omega_c$) and the other to a
suppressed value (say, around $\frac 1 2 \omega_c$), we predict
that the effect of the second source will be to suppress the peak
of the first source, even to destroy the peak completely if the
coherent effect is sufficiently strong.

\begin{acknowledgments}
We thank R. Chiao, R.R. Du, T.H. Hansson, and Z. Wang for helpful
discussions. DHL is supported by DOE DHL is supported by DOE grant
DE-AC03-76SF00098. JML has been associated with the Miller
Institute for Basic Research in Science as a visiting professor in
the fall of 2002 and will acknowledge the financial support and
the hospitality of the Miller institute.
\end{acknowledgments}

\widetext

\end{document}